\documentclass[12pt]{article}
\pdfoutput=1

\usepackage{amssymb,graphicx}
\usepackage[intlimits]{amsmath}
\usepackage{pst-all}
\usepackage{bbm}
\usepackage{slashed}
\usepackage[small]{subfigure}
\usepackage{MnSymbol}

\makeatletter \@addtoreset{equation}{section} \makeatother

\makeatletter
\let\old@startsection=\@startsection
\let\oldl@section=\l@section
\renewcommand{\@startsection}[6]{\old@startsection{#1}{#2}{#3}{#4}{#5}{#6\mathversion{bold}}}
\renewcommand{\l@section}[2]{\oldl@section{\mathversion{bold}#1}{#2}}
\makeatother

\makeatletter
\let\old@makecaption=\@makecaption
\def\@makecaption{\small\old@makecaption}
\makeatother

\def\x{{\tt x}}

\def\be{\begin{equation}}
\def\ee{\end{equation}}
\def\bee{\begin{eqnarray}}
\def\eee{\end{eqnarray}}
\def\nn{\nonumber}

\def\Str{\text{Str}}

\begin{document}

\begin{flushright}\footnotesize
\texttt{NORDITA-2012-67} \\
\texttt{UUITP-24/12}
\vspace{0.6cm}
\end{flushright}

\renewcommand{\thefootnote}{\fnsymbol{footnote}}
\setcounter{footnote}{0}

\begin{center}
{\Large\textbf{\mathversion{bold} B-field in $AdS_3/CFT_2$ Correspondence  \\
and Integrability}
\par}

\vspace{0.8cm}

\textrm{A.~Cagnazzo$^{1}$ and
K.~Zarembo$^{1,2}$\footnote{Also at ITEP, Moscow, Russia}}
\vspace{4mm}

\textit{${}^1$Nordita, KTH Royal Institute of Technology and Stockholm University, 
Roslagstullsbacken 23, SE-106 91 Stockholm, Sweden}\\
\textit{${}^2$Department of Physics and Astronomy, Uppsala University\\
SE-751 08 Uppsala, Sweden}\\
\vspace{0.2cm}
\texttt{cagnazzo@kth.se, zarembo@nordita.org}

\vspace{3mm}


\thispagestyle{empty}
\par\vspace{1cm}

\textbf{Abstract} \vspace{3mm}

\begin{minipage}{13cm}
We construct topological Wess-Zumino term for supercoset sigma-models on various $AdS_3$ backgrounds. For appropriately chosen set of parameters, the sigma-model remains integrable when the Wess-Zumino term is added to the action. Moreover, the conditions for integrability, kappa-symmetry and conformal invariance are equivalent to each other. 
\end{minipage}

\end{center}

\vspace{0.5cm}


\newpage
\setcounter{page}{1}
\renewcommand{\thefootnote}{\arabic{footnote}}
\setcounter{footnote}{0}


\section{Introduction}

The use of integrability in the AdS/CFT correspondence has made feasible non-perturbative calculations that would be too difficult or impossible otherwise \cite{Beisert:2010jr}. The prime example is the duality between $\mathcal{N}=4$ super-Yang-Mills theory in four dimensions and  type IIB string theory on $AdS_5\times S^5$ with the Ramond-Ramond (RR) flux, where the exact string spectrum, equivalently the spectrum of anomalous dimensions in the SYM theory can be described by Bethe-ansatz equations. Integrability methods are potentially applicable to other  $AdS_{d+1}$ backgrounds with RR flux and consequently to the dual $d$-dimensional CFTs.

We focus on the $d=2$ case, which  is special in many respects. For one thing, the three-form that prevents $AdS_3$ from collapsing to zero size can be an arbitrary combination of the RR and NSNS fluxes in contradistinction to  higher-dimensional, pure RR backgrounds. The RR backgrounds are often considered difficult to quantize. In this respect  $AdS_3$ is definitely simpler, since in the pure NSNS case the string can be quantized by more or less conventional methods of the worldsheet CFT \cite{Giveon:1998ns,Elitzur:1998mm,deBoer:1998pp,Maldacena:2000hw,Maldacena:2000kv,Maldacena:2001km}. The RR $AdS_3$ backgrounds are more complicated from the CFT perspective \cite{Berkovits:1999im,Ashok:2009jw}.
On the other hand they are integrable \cite{Chen:2005uj,Babichenko:2009dk}, and many results obtained for $AdS_5\times S^5$ can be transplanted to $AdS_3$ with minimal modifications. Along these lines, the algebraic curve \cite{Babichenko:2009dk}, the asymptotic Bethe equations \cite{Babichenko:2009dk,Sax:2011ms} and the Y-system \cite{Benichou:2010ts} have been constructed for strings in $AdS_3\times S^3\times T^4$ and $AdS_3\times S^3\times S^3\times S^1$.

A novel feature of the $AdS_3$ backgrounds is the presence of massless modes in the light-cone gauge, which cannot be straightforwardly included in the framework of  integrability. As a result, only a subset of the full string spectrum is currently known. Implications of the massless modes for integrability have been discussed recently \cite{Sax:2011ms,Rughoonauth:2012qd,Sundin:2012gc}, but they have not yet been fully incorporated in the Bethe-ansatz equations.

As far as mixed RR/NSNS $AdS_3$ backgrounds are concerned, very little is known, either from the CFT perspective or from the integrability point of view. Small deviations from the pure NSNS point are amenable to conformal perturbation theory \cite{Quella:2007sg}, although in the closed string sector an infinitesimal RR flux is actually a singular perturbation  \cite{Quella:2007sg}. We will approach the problem from the opposite direction, starting with the pure RR background. The NSNS flux should correspond to a topological Wess-Zumino (WZ) term \cite{Novikov:1982ei,Witten:1983tw,Witten:1983ar} in the string sigma-model action. 
The existence of the topological WZ term for the type of integrable sigma-models that we are going to consider was mentioned in \cite{Rahmfeld:1998zn}, but to the best of our knowledge, it has never been constructed explicitly.
There exists an alternative formulation of the GS string on $AdS_3\times S^3\times T^4$ \cite{Pesando:1998wm}, which accommodates both RR and NSNS fluxes, but integrability in this formulation is not really manifest.

The GS action on sufficiently symmetric RR backgrounds admits a supercoset formulation, as was first suggested by Metsaev and Tseytlin  for strings on  $AdS_5\times S^5$ \cite{Metsaev:1998it}. The key feature of this construction is a $\mathbbm{Z}_4$ symmetry \cite{Berkovits:1999zq}, which makes the supercoset, in the mathematics terminology, a semi-symmetric superspace \cite{Serganova}. The $\mathbbm{Z}_4$ symmetry guarantees that the classical equations of motion of the sigma-model have a Lax representation \cite{Bena:2003wd} and therefore possess an infinite set of integrals of motion. Imposing the conditions of conformal invariance and the central charge constraint on all possible semi-symmetric cosets \cite{Serganova,Adam:2007ws}
leaves a finite number of  $\mathbbm{Z}_4$ models  potentially consistent as string theories \cite{Zarembo:2010sg}.
Among them are two integrable $AdS_3$ backgrounds: $PSU(1,1|2)^2/SU(1,1)\times SU(2)={\rm Super}(AdS_3\times S^3)$ \cite{Rahmfeld:1998zn,Park:1998un,Metsaev:2000mv,Chen:2005uj,Babichenko:2009dk} and $D(2,1;\alpha )^2/SU(1,1)\times SU(2)^2={\rm Super}(AdS_3\times S^3\times S^3)$ \cite{Babichenko:2009dk}. 

It is clear from the outset that a generic $\mathbbm{Z}_4$ coset will not admit a WZ term because of the well-known geometric obstruction for gauging a symmetry subgroup in the WZ action \cite{Hull:1989jk,Witten:1991mm}.  The coset symmetry acts by right group multiplication in the $\mathbbm{Z}_4$ models, and this is exactly the case when the WZ action cannot be defined for the usual bosonic sigma-models \cite{Witten:1991mm}. The $AdS_3$ cosets, however, belong to a special class of semi-symmetric superspaces, whose bosonic section is a group manifold, for which the WZ action can  be easily constructed. We call these superspaces permutation cosets, because the underlying $\mathbbm{Z}_4$ symmetry acts on the Lie algebra of the symmetry group by semi-graded permutation. We will  construct the WZ action for an arbitrary coset of this type, starting with the WZ term on its bosonic section. The $AdS_3$ string backgrounds with the B-field switched on are just particular cases of this construction, when the global symmetry of the sigma model is $PSU(1,1|2)^2$ or $D(2,1;\alpha )^2$. We will then check if the sigma-model remains integrable, kappa-symmetric and conformal invariant after the WZ term is added to the action.

\section{Permutation supercosets}

A semi-symmetric superspace is a coset $\mathcal{G}/\mathcal{H}_0$ of a supergroup $\mathcal{G}$ over a bosonic subgroup $\mathcal{H}_0$, such that $\mathcal{H}_0$ stays invariant under the action of a $\mathbbm{Z}_4$ automorphism $\Omega $. The automorphism acts linearly on the Lie algebra of $\mathcal{G}$ and its fourth power is the identity: $\Omega ^4=\mathop{\mathrm{id}}$. The current $J$ of the sigma-model can be decomposed into components with definite $\mathbbm{Z}_4$ charge:
\begin{equation}\label{Z4action}
 \Omega (J_n)=i^nJ_n.
\end{equation}
The action of the supercoset is simply \cite{Berkovits:1999zq}
\begin{equation}
\label{sigma_RR}
 S_{\rm MT} =\frac{1}{2}\int_{\mathcal{M}}^{}\mathop{\mathrm{Str}}\left(J_2\wedge *J_2+J_1\wedge J_3\right).
\end{equation}
For various supergravity backgrounds that contain RR fields, this construction gives the GS string action, sometimes after partially fixing the kappa-symmetry gauge. 

We will be interested in a particular class of semi-symmetric cosets, in which the symmetry group is the direct product of two simple supergroups: $\mathcal{G}=G\times G$. As noticed in \cite{Babichenko:2009dk}, a direct sum of two superalgebras always admits a $\mathbbm{Z}_4$ action, defined as a semi-graded permutation of the two factors:
\begin{equation}\label{Omega}
 \Omega =\begin{pmatrix}
  0 & \mathop{\mathrm{id}} \\
  \left(-1\right)^F & 0 \\
 \end{pmatrix}.
\end{equation}
The invariant subspace of this automorphism is the diagonal bo\-so\-nic subalgebra of $\mathfrak{g}\oplus\mathfrak{g}$: the set of elements of the form $(\xi ,\xi )$, $\xi \in \mathfrak{g}^B$, where $\mathfrak{g}$ is the Lie algebra of $G$ and $\mathfrak{g}^B$ is its Grassmann-even subalgebra.

The supercoset then has the form $G\times G/G^B_{\rm diag}$. Its bosonic section,  $G^B\times G^B/G^B_{\rm diag}=G^B$,  is just the group manifold of the bosonic subgroup of $G$. If we take $G=PSU(1,1|2)$, its bosonic subgroup is $SU(1,1)\times SU(2)$, which as a manifold is isomorphic to the direct product $AdS_3\times S^3$. The odd embedding coordinates of the supercoset, that arise from the sixteen supercharges of $\mathfrak{psu}(1,1|2)\oplus\mathfrak{psu}(1,1|2)$, can be interpreted as the GS fermions which remain after fixing the kappa-symmetry gauge in the GS action on $AdS_3\times S^3\times T^4$. Supplementing the coset action with four flat bosonic coordinates yields the GS action on  $AdS_3\times S^3\times T^4$ in a particular kappa-symmetry gauge \cite{Babichenko:2009dk}. The same construction for $G=D(2,1;\alpha )$ requires one additional boson and yields the GS action on $AdS_3\times S^3\times S^3\times S^1$, since the even subgroup of an appropriate real form of $D(2,1;\alpha )$ is $SU(1,1)\times SU(2)\times SU(2)$.

Let us detail how the $\mathbbm{Z}_4$ construction works for the permutation supercosets. The string embedding coordinates are parameterized by a pair of supergroup elements: $(g_L(\sigma ),g_R(\sigma ))$, $g_{L,R}\in G$, subject to gauge transformations $g_{L,R}\rightarrow g_{L,R}h$, where the same $h\in G^B$ acts on the two coset representatives simultaneously. The global $G\times G$ symmetry acts by independent multiplications from the left: $g_{L,R}\rightarrow h_{L,R}g_{L,R}$.

The action and the equations of motion of the sigma-model can be written in terms of the left-invariant currents:
\begin{equation}
 J_{L,R}=g_{L,R}^{-1}dg_{L,R}.
\end{equation}
To define the $\mathbbm{Z}_4$  action,
we first decompose the currents into the bosonic (even) and fermionic (odd) components, according to the superalgebra's Grassmann parity:
\begin{equation}
 J_{L,R}=J_{L,R}^B+J_{L,R}^F,
\end{equation}
The $\mathbbm{Z}_4$ automorphism then acts according to (\ref{Omega}):
\begin{equation}
 \Omega (J_{L,R}^B)=J_{R,L}^B,\qquad \Omega (J_{L,R}^F)=\mp J_{R,L}^F.
\end{equation}
The $\mathbbm{Z}_4$ decomposition, consistent with (\ref{Z4action}), is given by
\begin{eqnarray}\label{Z4dec}
 J_0&=&\frac{1}{2}\left(J_L^B+J_R^B\right)
\nonumber \\
J_1&=&\frac{1}{2}\left(J_L^F+iJ_R^F\right)
\nonumber \\
 J_2&=&\frac{1}{2}\left(J_L^B-J_R^B\right)
\nonumber \\
J_3&=&\frac{1}{2}\left(J_L^F-iJ_R^F\right).
\end{eqnarray}
The sigma-model action is then defined by (\ref{sigma_RR}).

The gauge symmetry acts on the currents as
\begin{equation}
 J_{L,R}^B\rightarrow h^{-1}J_{L,R}^Bh+h^{-1}dh,\qquad
  J_{L,R}^F\rightarrow h^{-1}J_{L,R}^Fh.
\end{equation}
Since the $h^{-1}dh$ term cancels in $J_2$, the action (\ref{sigma_RR}) is manifestly gauge-invariant. It is also invariant under $\mathbbm{Z}_4$ transformations. The current
$J_0$ transforms under gauge transformations as a connection, and thus plays the r\^ole of a non-dynamical gauge field for the coset symmetry.

As a consequence of their definition,
the currents satisfy the Maurer-Cartan equations, which can also be projected onto even and odd subspaces in $\mathfrak{g}$:
\begin{eqnarray}\label{Maurer-Cartan}
 dJ_{L,R}^B+J_{L,R}^B\wedge J_{L,R}^B+J_{L,R}^F\wedge J_{L,R}^F&=&0
\nonumber \\
 dJ_{L,R}^F+J_{L,R}^B\wedge J_{L,R}^F+J_{L,R}^F\wedge J_{L,R}^B&=&0.
\end{eqnarray}
The Maurer-Cartan equations can be written in a manifestly gauge-invariant form by introducing the field strength of $J_0$:
\begin{equation}
 F=dJ_0+J_0\wedge J_0,
\end{equation}
and the covariant exterior derivative, that acts on any $\mathfrak{g}$-valued $p$-form $C_p$ according to
\begin{equation}\label{D_def}
 DC_p=dC_p+J_0\wedge C_p+\left(-1\right)^{p+1}C_p\wedge J_0.
\end{equation}
Taking particular linear combinations of the four equations in (\ref{Maurer-Cartan}), we arrive at the manifestly gauge-invariant form of the Maurer-Cartan equations:
\begin{eqnarray}\label{Maurer-C}
 F+J_2\wedge J_2+J_1\wedge J_3+J_3\wedge J_1&=&0
\nonumber \\
DJ_2+J_1\wedge J_1+J_3\wedge J_3&=&0
\nonumber \\
DJ_1+J_2\wedge J_3+J_3\wedge J_2&=&0
\nonumber \\
DJ_3+J_2\wedge J_1+J_1\wedge J_2&=&0.
\end{eqnarray}

\section{WZ term}

The WZ term is an integral over a three-dimensional ball $\mathcal{B}$ whose boundary is the string worldsheet: $\partial \mathcal{B}=\mathcal{M}$. The integrand must locally be a total derivative, such that the variation of the WZ action integrates to a two-dimensional expression, yielding the equations of motion that only depend on the fields on $\mathcal{M}$ \cite{Novikov:1982ei}.

The bosonic part of a permutation coset is just the sigma-model on the group manifold of $G^B$, for which the WZ term has the standard form of the wedge product of three currents integrated over $\mathcal{B}$ \cite{Witten:1983ar}. The only current that survives the bosonic truncation is $J_2$, so the first guess for how the WZ term (for a $\mathbbm{Z}_4$ coset) could look like is
\begin{equation}\label{WZbare}
 S^{\rm bos}_{\rm WZ}=\frac{2}{3 }\int_{\mathcal{B}}^{}
 \mathop{\mathrm{Str}}J_2\wedge J_2\wedge J_2.
\end{equation}
This expression, however, cannot be the full answer. We will shortly demonstrate that the variation of the integrand is not a total derivative. We need to supplement this action with extra  terms, which cancel the three-dimensional  part of its variation and
make the equations of motion consistently two-dimensional. The requisite fermionic completion, as we shall see, does exist and is essentially unique.

Under infinitesimal variations of the fields $\delta g_{L,R}=g_{L,R}\xi _{L,R}$, $\xi _{L,R}\in\mathfrak{g}$, the currents transform as
\begin{equation}
 \delta J_{L,R}=d\xi _{L,R}+[J_{L,R},\xi _{L,R}].
\end{equation}
The variations of their $\mathbbm{Z}_4$ components are
\begin{eqnarray}\label{deltacurrents}
 \delta J_0&=&D\xi _0+[J_2,\xi _2]+[J_1,\xi _3]+[J_3,\xi _1]
\nonumber \\
\delta J_2&=&D\xi _2+[J_2,\xi _0]+[J_1,\xi _1]+[J_3,\xi _3]
\nonumber \\
\delta J_1&=&D\xi _1+[J_1,\xi _0]+[J_2,\xi _3]+[J_3,\xi _2]
\nonumber \\
\delta J_3&=&D\xi _3+[J_3,\xi _0]+[J_2,\xi _1]+[J_1,\xi _2],
\end{eqnarray}
where $\xi _n$ are the $\mathbbm{Z}_4$ projections of the variation parameter, which are defined similarly to (\ref{Z4dec}) and which satisfy eq.~(\ref{Z4action}). The zero-grading component $\xi _0$ is a parameter of an infinitesimal gauge transformation, and should drop from the variation of the action in virtue of  the gauge invariance.

Taking the variation of  (\ref{WZbare}), we find:
 \bee\label{varWZ}
\delta S^{\rm bos}_{\rm WZ}
&=&2\int_\mathcal{B}\,\Str[d(\xi_2 J_2\wedge J_2)+\xi_2(J_1\wedge J_1\wedge J_2-J_2\wedge J_1\wedge J_1
\nonumber \\
&&+J_3\wedge J_3\wedge J_2-J_2\wedge J_3\wedge J_3)-\xi_1 (J_1\wedge J_2\wedge J_2-J_2\wedge J_2\wedge J_1)
\nonumber \\
&&-\xi_3(J_3\wedge J_2\wedge J_2-J_2\wedge J_2\wedge J_3)].
\eee
In simplifying this expression we used the Maurer-Cartan equations. The first term is what we would have gotten for the purely bosonic model on a group manifold. The other terms include fermionic currents and do not combine to total derivatives. To cancel the non-locality in the action's variation we need to add extra fermionic terms.

Additional terms should have grading two, and should be symmetric under the interchange of $J_1$ and $J_3$, whose labeling is a matter of convention, changed by taking $\Omega ^3$ as a generator of $\mathbbm{Z}_4$. An integral that satisfies these condition is actually unique:
\bee\label{defI}
I=\int_\mathcal{B}\, \Str(J_3\wedge J_1\wedge J_2+J_1\wedge J_3\wedge J_2).
\eee
Its variation is
\bee
\delta I&=&\int_\mathcal{B}\,\Str\{\,d[\xi_2 (J_1\wedge J_3+ J_3\wedge J_1)+\xi_1 (J_2\wedge J_3+J_3\wedge J_2)
\nonumber \\
&&
+\xi_3 (J_2\wedge J_1+J_1\wedge J_2)]\nn\\
&&-2\xi_2(J_1\wedge J_1\wedge J_2-J_2\wedge J_1\wedge J_1+J_3\wedge J_3\wedge J_2-J_2\wedge J_3\wedge J_3)\nn\\
&&+2\xi_1 (J_1\wedge J_2\wedge J_2-J_2\wedge J_2\wedge J_1)+2\xi_3(J_3\wedge J_2\wedge J_2-J_2\wedge J_2\wedge J_3)\}.\nn\\
\eee
The last two lines have exactly the right form to cancel the non-local part in (\ref{varWZ}). We thus should add  (\ref{defI}) to (\ref{WZbare}) with the coefficient one. The resulting action,
\begin{equation}\label{WZW_grad2}
  S_{\rm WZ}=\int_{\mathcal{B}}^{}
 \mathop{\mathrm{Str}}\left(
 \frac{2}{3}\, J_2\wedge J_2\wedge J_2
 +J_1\wedge J_3\wedge J_2+
 J_3\wedge J_1\wedge J_2
 \right).
\end{equation}
has a local variation:
\begin{eqnarray}\label{variationWZ}
 \delta S_{\rm WZ}&=&\int_{\mathcal{M}}
 \mathop{\mathrm{Str}}
 \left[\xi _2\left(2J_2\wedge J_2+J_1\wedge J_3+J_3\wedge J_1\right)
 \right.
\nonumber \\
&&\left.+\xi _1\left(J_2\wedge J_3+J_3\wedge J_2\right)
 +\xi _3\left(J_2\wedge J_1+J_1\wedge J_2\right)\right].
\end{eqnarray}

This is the unique topological WZ term that can be added to the action of any permutation coset.  One can find another cubic combination of currents whose variation is a total derivative. This combination has grading zero, and in fact can be explicitly written  as a total derivative, after which the corresponding WZ term integrates to $J_1\wedge J_3$, which is nothing but the GS term in the sigma-model action (see \cite{Berkovits:1999zq} for more details).

It might seem that we have not used any special properties of  permutation cosets, since the derivation relied solely on the Maurer-Cartan equations (\ref{Maurer-C}) and the $\mathbbm{Z}_4$ structure of the variations (\ref{deltacurrents}). Both are the same for any semi-symmetric supercoset. However,
for cosets based on simple supergroups, the WZ term (\ref{WZW_grad2}) will merely vanish, by $\mathbbm{Z}_4$ invariance of the supertrace. A peculiar feature of the permutation supercosets is that grading two and grading zero subspaces are not orthogonal, which ultimately allowed us to construct a WZ term for this class of sigma-models.

\section{Equations of motion and integrability}

We now consider the action that contains all three terms discussed above: the sigma-model term, the GS term and the WZ term, with arbitrary relative coefficients.  An arbitrary coefficient for the GS term, that was fixed to one in the action without the B-field, eq.~(\ref{sigma_RR}), is necessary for integrability and kappa-symmetry.

Our  starting point is thus
\begin{eqnarray}\label{mainaction}
 S&=&\frac{1}{2}\int_{\mathcal{M}}^{}\mathop{\mathrm{Str}}
 \left(J_2\wedge *J_2+\kappa J_1\wedge J_3\right)
\nonumber \\
&&+
 \chi \int_{\mathcal{B}}^{}
 \mathop{\mathrm{Str}}\left(\frac{2}{3}\,J_2\wedge J_2\wedge J_2
 +J_1\wedge J_3\wedge J_2+
 J_3\wedge J_1\wedge J_2\right),
\end{eqnarray}
where $\kappa $ and $\chi $ are so far arbitrary coupling constants. This action is no longer $\mathbbm{Z}_4$-invariant, because the WZ term has an overall grading two. Our goal is to check if the kappa-symmetry, the conformal invariance and the integrability are preserved at non-zero $\chi $. We start with integrability.

Using (\ref{deltacurrents}) for the variation of the currents and taking  the variation of the WZ term from (\ref{variationWZ}), we get the following equations of motion:
\bee\label{eqs_of_motion}
D*J_2-\kappa \,J_1\wedge J_1+\kappa \,J_3\wedge J_3- 2\chi J_2\wedge J_2-\chi J_1\wedge J_3-\chi J_3\wedge J_1 &=&0\nn\\
(\kappa J_1+*J_1)\wedge J_2+J_2\wedge(\kappa J_1+*J_1)+\chi \left(J_2\wedge J_3+J_3\wedge J_2\right)&=&0\nn\\
(\kappa J_3-*J_3)\wedge J_2+J_2\wedge(\kappa J_3-*J_3)-\chi \left(J_2\wedge J_1+J_1\wedge J_2\right)&=&0\nn\\
\eee
If $\chi =0$, these equations admit a Lax representation \cite{Bena:2003wd}, which then guarantees the existence of an infinite number of conserved charges, making the model classically integrable. We would like to formulate the conditions under which this Lax connection can be  deformed to include the WZ coupling.

 To this end, we take the following ansatz:
\be
L=J_0+\alpha_1J_2+\alpha_2 *J_2+\beta_1J_1+\beta_2 J_3.
\ee
The model is integrable if the flatness of the Lax connection is equivalent to the full set of the equations of motion, including  the Maurer-Cartan equations (\ref{Maurer-C}). We thus require that
\be\label{flatnessL}
dL+L\wedge L=0,
\ee
provided that (\ref{Maurer-C}) and (\ref{eqs_of_motion}) are satisfied.
This leads to the following overconstrained system of equations on the coefficients $\alpha _i$, $\beta _i$:
\bee
-\alpha_1+\kappa \,\alpha_2+\beta_1^2&=&0\nn\\
-\alpha_1-\kappa \,\alpha_2+\beta_2^2&=&0\nn\\
\chi \alpha_2-1+\beta_1\beta_2&=&0\nn\\
2\chi  \alpha_2-1+\alpha_1^2-\alpha_2^2&=&0\nn\\
-\beta_1+\alpha_1\beta_2+\chi \alpha_2\beta_1-\kappa \,\alpha_2\beta_2&=&0\nn\\
-\beta_2+\alpha_1\beta_1+\chi \alpha_2\beta_1+\kappa \,\alpha_2\beta_1&=&0.
\eee
These equations have no solutions, unless  the parameters $\kappa $ and $\chi $ are related:
\be\label{kappathroughchi}
\kappa ^2=1-\chi ^2.
\ee
The equations on $\alpha _i$, $\beta _i$ then become underconstrained, and have a one-parametric set of solutions:
\bee\label{alphabeta}
\alpha_2 &=& \chi \pm \sqrt{-1 + \alpha_1^2 + \chi ^2}\nn\\
\beta_1&=&\pm\sqrt{\alpha_1 -\kappa \,\alpha_2}\nn\\
\beta_2&=&\pm\sqrt{\alpha_1 +\kappa \,\alpha_2}\,.
\eee
One of the unknowns (here $\alpha _1$) is not fixed by the equations and plays the r\^ole of the spectral parameter.

 For many purposes a different parameterization of the Lax connection is more convenient. It is desirable to introduce the spectral parameter in such a way that the coefficients $\alpha _i$ and $\beta _i^2$ are rational functions. This is useful, for example, in the construction of the algebraic curve for the classical solutions of the sigma-model \cite{Beisert:2005bm}. To arrive at such a parameterization we can take
\begin{equation}
 \alpha _1=\kappa \,\frac{\x^2+1}{\x^2-1}\,.
\end{equation}
Then for the Lax connection we get:
\begin{eqnarray}
 L&=&J_0+\kappa \,\frac{\x^2+1}{\x^2-1}\,J_2+\left(\chi -\frac{2\kappa \x}{\x^2-1}\right)*J_2
\nonumber \\
&&
 +\left(\x+\frac{\kappa }{1-\chi }\right)\sqrt{\frac{\kappa \left(1-\chi \right)}{\x^2-1}}J_1
 +\left(\x-\frac{\kappa }{1+\chi }\right)\sqrt{\frac{\kappa \left(1+\chi \right)}{\x^2-1}}J_3.
\end{eqnarray}
This form of the Lax connection reduces to the standard one at $\chi =0$, $\kappa =1$. Working backwards one can easily see that both the Maurer-Cartan equations and the equations of motion follow from the flatness condition, if the latter holds for any value of the spectral parameter.

At $\chi =1$, $\kappa =0$, the Lax connection degenerates. This corresponds to the WZW point, where  a simpler condition of chiral (holomorphic) factorization for the currents replaces integrability.

\section{Background field method}\label{sec:background}

To study the properties of the two-dimensional field theory defined by  (\ref{mainaction}) we expand the action around a classical field configuration $(\bar{g}_L,\bar{g}_R)$ to the quadratic order in fluctuations. This will allow us to compute the one-loop beta-function for the sigma-model coupling, and also to find the mass spectrum of the string fluctuations in the light-cone gauge.

We assume (for simplicity) that the background fields are bosonic: $\bar{g}_{L,R}\in G^B$, and on-shell, so that the background currents $\bar{J}_{L,R}=\bar{g}_L^{-1}d\bar{g}_{L,R}$ satisfy the equations of motion. It is convenient to introduce special notations for their grading zero ($A$) and grading two ($K$) projections: $\bar{J}_{L,R}=A\pm K$. The currents satisfy the Maurer-Cartan equations:
\begin{eqnarray}\label{Cartan_backgr}
 F+K\wedge K&=&0
\nonumber \\
DK&=&0,
\end{eqnarray}
which are just identities that follow from definitions.
Here $F=dA+A\wedge A$, and the covariant derivative is defined as in (\ref{D_def}), with $J_0$ replaced by $A$. In addition, we assume that the currents satisfy the equations of motion:
\begin{equation}\label{eqm_backgr}
 D*K-2\chi K\wedge K=0.
\end{equation}
The equations above are obtained from (\ref{Maurer-C}) and (\ref{eqs_of_motion}) by setting fermion currents to zero.

Our goal is to expand the action to the second order in fluctuations, where the fluctuating fields $X_{L,R}$ are defined via
\begin{equation}
 g_{L,R}=\bar{g}_{L,R}\,{\rm e}\,^{X_{L,R}}.
\end{equation}
The background-field expansion of currents can be derived from the general formula
\begin{equation}\label{backgroundfieldexp}
 J=\bar{J}+\frac{1-\,{\rm e}\,^{-\mathop{\mathrm{ad}}X}}{{\mathop{\mathrm{ad}}X}}\,\mathcal{D}X
 =\bar{J}+\mathcal{D}X-\frac{1}{2}\,\left[X,\mathcal{D}X\right]+\ldots ,
\end{equation}
where
\begin{equation}
 \mathcal{D}X=dX+[\bar{J},X].
\end{equation}
Here $J$, $\bar{J}$, $\mathcal{D}$ and $X$ are  $J_{L,R}$, $\bar{J}_{L,R}$, $\mathcal{D}_{L,R}$ and $X_{L,R}$. For future convenience we introduce the covariant derivative twisted by a $\mathfrak{g}$-valued one-form $\omega $:
\begin{equation}
 {D}_{\omega} C_p=DC_p+\omega \wedge C_p+\left(-1\right)^{p+1}C_p\wedge \omega .
\end{equation}
The left and right derivatives, which appear in the background-field expansion of the currents, can then be written as
\begin{equation}
 \mathcal{D}_{L,R}={D}_{\pm K}.
\end{equation}

Not all the components of the fluctuation fields $X_{L,R}$ are independent dynamical variables, as they are subject to the coset gauge transformations. It is convenient to fix the gauge freedom from the outset, prior to expanding the action. We impose $X^B_L=-X^B_R$ as a gauge condition. In terms of the $\mathbbm{Z}_4$ components of $X$, $X_p$ with $p=0\ldots 3$, this is equivalent to setting
\begin{equation}
 X_0=0.
\end{equation}
The remaining components describe bosonic ($X_2$) and fermionic ($X_1$ and $X_3$) degrees of freedom of the superstring. Under simultaneous gauge transformations that act both on the full quantum fields of the sigma-model and on the background, the fluctuations $X_p$ ($p=1,2,3$) transform in the adjoint: $X_p\rightarrow h^{-1}X_ph$.

Once the coset gauge is fixed, we can expand the currents according to (\ref{backgroundfieldexp}), decompose the fluctuation fields into their $\mathbbm{Z}_4$ components and substitute the result in the action of the sigma-model. Along the way we will need the background-field expansion of the $\mathbbm{Z}_4$ currents:
\begin{eqnarray}
 J_2&=&K+DX_2-\frac{1}{2}\,\left[X_2,\left[K,X_2\right]\right]
-\frac{1}{2}\,\left[X_1,DX_1\right]
  -\frac{1}{2}\,\left[X_3,DX_3\right]\nonumber \\
&&
  -\frac{1}{2}\,\left[X_1,\left[K,X_3\right]\right]
    -\frac{1}{2}\,\left[X_3,\left[K,X_1\right]\right]
    +O\left(X^3\right)
\nonumber \\
J_1&=&DX_1+\left[K,X_3\right]+O\left(X^2\right)
\nonumber \\
J_3&=&DX_3+\left[K,X_1\right]+O\left(X^2\right).
\end{eqnarray}
Expanding the action to the quadratic order in $X$ is then a straightforward albeit a lengthy exercise. 

The following identities turn out to be useful in bringing the WZ term to the local 2d form:
\begin{eqnarray}
 D^2C_p&=&C_p\wedge K\wedge K-K\wedge K\wedge C_p
\nonumber \\
d\mathop{\mathrm{Str}}\left[K,X\right]\wedge DY
&=&\mathop{\mathrm{Str}}\left(
K\wedge K\wedge \left[Y,\left[K,X\right]\right]\right.
\nonumber \\
&&\left.
-K\wedge DX\wedge DY-K\wedge DY\wedge DX
\right).
\end{eqnarray}
These identities follow from the Maurer-Cartan equations for the background currents, and do not depend on the equations of motion. Using these equations,  the WZ term can be integrated and brought to a manifestly two-dimensional form:
\begin{eqnarray}\label{S2}
 S^{(2)}&=&\frac{1}{2}\int_{}^{}\mathop{\mathrm{Str}}
 \left(
 DX_2\wedge *DX_2-\left[K,X_2\right]\wedge *\left[K,X_2\right]
 +2\chi DX_2\wedge \left[K,X_2\right]\right.
\nonumber \\
&&\left.
 +X_1D*\left[K,X_1\right]
 -\kappa X_1D\left[K,X_1\right]
 +X_3D*\left[K,X_3\right]
 +\kappa X_3D\left[K,X_3\right]
 \right.
\nonumber \\
&&\left.
 -\chi X_1D\left[K,X_3\right]
 -\chi X_3D\left[K,X_1\right]
  \right.
\nonumber \\
&&\left.
 -2\left[K,X_1\right]\wedge *\left[K,X_3\right]
  -2\kappa \left[K,X_1\right]\wedge \left[K,X_3\right]
 \right).
\end{eqnarray}

The bosonic part of the action can be compactly written as
\begin{equation}
\label{quadbosonic}
 S_B^{(2)}=\frac{1}{2}\int_{}^{}\mathop{\mathrm{Str}}
 \left\{
 D_{\chi *K}X_2\wedge *D_{\chi *K}X_2
 -\left(1-\chi ^2\right)\left[K,X_2\right]\wedge *\left[K,X_2\right]
 \right\}.
\end{equation}
The concise form of the fermion action is
\begin{equation}\label{quadferm}
 S_F^{(2)}=\frac{1}{2}\int_{}^{}\mathop{\mathrm{Str}}
 X_I
 \left(D+\sigma _1\mathop{\mathrm{ad}}K\wedge \right)^{IJ}
 \left(*-\kappa \sigma _3-\chi \sigma _1\right)^{JL}
 \mathop{\mathrm{ad}}K\,X_L,
\end{equation}
where the $\mathbbm{Z}_4$ indices $I$, $J$ and $L$ take values $1$ or $3$ and are carried by the Pauli matrices. The summation over repeated indices is implied. All operators, like the Hodge $*$ and $\mathop{\mathrm{ad}}K$ act on everything to their right. For instance, $\left(*-\kappa \sigma _3-\chi \sigma _1\right)^{JL}
 \mathop{\mathrm{ad}}K\,X_L$ is a shorthand notation for
 $[*K,X_J]-\kappa \sigma _3^{JL}[K,X_L]-\chi \sigma _1^{JL}[K,X_L]$.

The Lagrangian in (\ref{quadferm}) differs from the fermion Lagrangian in (\ref{S2}) by a term of the form $\chi \mathop{\mathrm{Str}}[K,X_I]\wedge \left[K,X_I\right]$. This term equals to zero, because of the anti-symmetry of the wedge product and the cyclic symmetry of the supertrace. We have added this term deliberately, to make the Dirac operator manifestly Hermitean.

To check the Hermiticity, we can use the following identities:
\begin{eqnarray}
 \mathop{\mathrm{ad}}K\wedge D&=&-D\mathop{\mathrm{ad}}K
\nonumber \\
\label{strange_identity}
*\mathop{\mathrm{ad}}K\wedge D&=&-D*\mathop{\mathrm{ad}}K
+2\chi \mathop{\mathrm{ad}}K\wedge \mathop{\mathrm{ad}}K,
\end{eqnarray}
which follow the Maurer-Cartan equations (\ref{Cartan_backgr}), as well as from the equations of motion (\ref{eqm_backgr}).  It is understood that the derivatives act on everything to their right. Taking the Hermitean conjugate of the fermion quadratic form, we find:
\begin{eqnarray}
 &&\left[
 \left(D+\sigma _1\mathop{\mathrm{ad}}K\wedge \right)
 \left(*-\kappa \sigma _3-\chi \sigma _1\right)\mathop{\mathrm{ad}}K
 \right]^\dagger
\nonumber \\
&&
 =
 -\left(*-\kappa \sigma _3-\chi \sigma _1\right)\mathop{\mathrm{ad}}K
 \wedge \left(D+\sigma _1\mathop{\mathrm{ad}}K\right)
\nonumber \\
&&
 =
 \left(D+\sigma _1\mathop{\mathrm{ad}}K\wedge \right)
 \left(*-\kappa \sigma _3-\chi \sigma _1\right)\mathop{\mathrm{ad}}K
\nonumber \\
&&
 -2\chi  \mathop{\mathrm{ad}}K\wedge \mathop{\mathrm{ad}}K
 +2\sigma _1\left(\kappa \sigma _3+\chi \sigma _1\right)\mathop{\mathrm{ad}}K\wedge \mathop{\mathrm{ad}}K
 -\kappa \left[\sigma _3,\sigma _1\right]\mathop{\mathrm{ad}}K\wedge \mathop{\mathrm{ad}}K
\nonumber \\
&& \nonumber
 =
  \left(D+\sigma _1\mathop{\mathrm{ad}}K\wedge \right)
 \left(*-\kappa \sigma _3-\chi \sigma _1\right)\mathop{\mathrm{ad}}K.
\end{eqnarray}
 The Hermiticity of the Dirac operator is the only place where we have used the equations of motion. The derivation of the action  (\ref{S2}) relied only on the Maurer-Cartan equations, which are just kinematic identities.

\section{Kappa-symmetry}

An important property of the GS action is the kappa-symmetry, a local fermionic symmetry that allows one to gauge away the unphysical components of the world-sheet fermions. In the semi-symmetric cosets the kappa-symmetry is related to the algebraic structure of the $\mathbbm{Z}_4$ decomposition of the underlying superalgebra: $\mathfrak{g}\oplus\mathfrak{g}=\mathfrak{h}_0\oplus\mathfrak{h}_1\oplus\mathfrak{h}_2\oplus\mathfrak{h}_3$, where elements of $\mathfrak{h}_n$ have $\mathbbm{Z}_4$ charge $n$. For  $\mathbbm{Z}_4$ cosets without the WZ term, the rank of the kappa-symmetry is determined by the commutant of two fixed, but sufficiently generic elements of $\mathfrak{h}_2$ \cite{Zarembo:2010sg,Vicedo:2010qd}:
\begin{equation}\label{rankkappa}
 \mathop{\mathrm{rank}}\nolimits_\kappa=\left.\dim\ker \mathop{\mathrm{ad}}K_+\right|_{\mathfrak{h}_1}
 +\left.\dim\ker \mathop{\mathrm{ad}}K_-\right|_{\mathfrak{h}_3}.
\end{equation}
To the second-order in the background-field expansion, the kappa-symmetry acts as linear shifts of $X_1$ and $X_3$ that commute with the light-cone components of the background current $K$. 

We may expect that one of the consistency requirements for the string propagation in  a B-field is the kappa-symmetry of the string action. The total rank of the kappa-symmetry should be independ of the WZ coupling. The transformation rules, on the contrary, may be deformed by the presence of the NSNS flux. 

To derived the conditions for unbroken kappa-symmetry, we need few extra definitions.
The chiral (light-cone) projection of a one-form is defined as
\begin{equation}
 \omega _\pm=\frac{1\mp *}{2}\,\omega .
\end{equation}
 The chiral components of the background current $K_\pm$ have only one component each. If we impose the Virasoro constraints, $K_\pm$ will in addition have a null supertrace norm. The rank of the kappa-symmetry ultimately depends on the commutation relations of the superalgebra and on whether we impose the Virasoro constraints or not.  
 
 Let us assume that the rank of the kappa-symmetry is different from zero, which according to (\ref{rankkappa}) means that the equations\begin{equation}
 \left[K_\pm,\epsilon ^\pm\right]=0
\end{equation}
have non-trivial solutions in the fermionic sugalgebra $\mathfrak{g}^F$. 
These equations can  be re-written as
\begin{equation}\label{*ad}
 *\mathop{\mathrm{ad}}K\,\epsilon ^\pm=\pm\mathop{\mathrm{ad}}K\,\epsilon
\end{equation}
The number of linearly independent solutions determines the rank of the kappa-symmetry at $\chi =0$, $\kappa =1$. We are going to check if the kappa-symmetry survives at non-zero $\chi $.

To this end, we will look for the shift symmetries of the action (\ref{quadferm}) of the form
\begin{equation}\label{kappa}
 \delta X_I=C_I^\pm\epsilon ^\pm,
\end{equation}
where $C_I^\pm$ are numerical constants.
Applying this transformation to the fermion action (\ref{quadferm}), and using (\ref{*ad}), we get\footnote{Here we used the Hermiticity of the Dirac operator by  varying only one $X$ and multiplying the result by two.}:
\begin{equation}
 \delta S_F^{(2)}=\int_{}^{}\mathop{\mathrm{Str}}
 X\left(D+\sigma _1\mathop{\mathrm{ad}}K\wedge \right)
 \left(\pm 1-\kappa \sigma _3-\chi \sigma _1\right)C^\pm
 \mathop{\mathrm{ad}}K\,\epsilon ^\pm.
\end{equation}
The variation vanishes if
\begin{equation}
\label{kappa_condition}
 \left(\pm 1-\kappa \sigma _3-\chi \sigma _1\right)C^\pm=0.
\end{equation}
These equations have a solution if and only if
\begin{equation}
 \det\left(\pm 1-\kappa \sigma _3-\chi \sigma _1\right)=
 1-\kappa ^2-\chi ^2=0.
\end{equation}

We got the same relationship between the couplings that guarantees   integrability of the sigma-model! The conditions for integrability and kappa-symmetry are thus equivalent. It is sufficient to require integrability, kappa-symmetry will then follow, or vice versa, it is enough to impose the kappa-symmetry, the sigma-model will then be automatically integrable.

We only considered linearized kappa-symmetry transformations (\ref{kappa}). It should be possible to uplift the kappa-symmetry of the full non-linear action (\ref{mainaction}). We will not do it here, as our goal was just to demonstrate the relationship between kappa-symmetry and integrability. The linearized form of the kappa-symmetry transformations will be also sufficient for computing the beta-function in the one-loop approximation.

\section{Beta Function}

We want to show that conformality is not spoiled by the introduction of the WZ term, in particular for the two supercosets we are interested in, i.e. $PSU(1,1|2)^2/SU(1,1)\times SU(2)$ and $D(2,1;\alpha )^2/SU(1,1)\times SU(2)^2$.
From now on we will assume the relation \eqref{kappathroughchi} between the couplings, since, as we have seen, it is crucial for integrability and kappa-symmetry.

Quantum-mechanically, the sigma-model is defined by a path integral
\begin{equation}\label{path_int}
 Z=\int_{}^{}D\phi \,{\rm e}\,^{\frac{i\sqrt{\lambda }S}{2\pi }},
\end{equation}
where $S$ is the action (\ref{mainaction}), the integration variables include $g_{L}$, $g_R$ and, in principle, the two-dimensional metric. The sigma-model coupling is denoted by $2\pi /\sqrt{\lambda }$. This is the standard convention in the AdS/CFT correspondence, where $\lambda $  is usually related to the 't~Hooft coupling of the dual 2d CFT. The gauge conditions for the coset gauge invariance and for the kappa-symmetry are of the unitary gauge type, and consequently there are no associated ghosts.

The zeroth-order consistency requirement for the path integral (\ref{path_int}) is the absence of the coupling constant renormalization.  We will compute the one-loop beta-function of $2\pi /\sqrt{\lambda }$ by substituting the background-field expansion (\ref{S2}) (equivalently (\ref{quadbosonic}), (\ref{quadferm})) into the path integral and integrating out the fluctuations, as done in many related work \cite{Polyakov:1975rr,Berkovits:1999zq,Polyakov:2004br,Vallilo:2002mh,Kagan:2005wt,Puletti:2006vb,Adam:2007ws,Mazzucato:2009fv,Zarembo:2010sg}. 

\begin{figure}[t]
\begin{center}
\subfigure[]{
   \includegraphics[height=2cm] {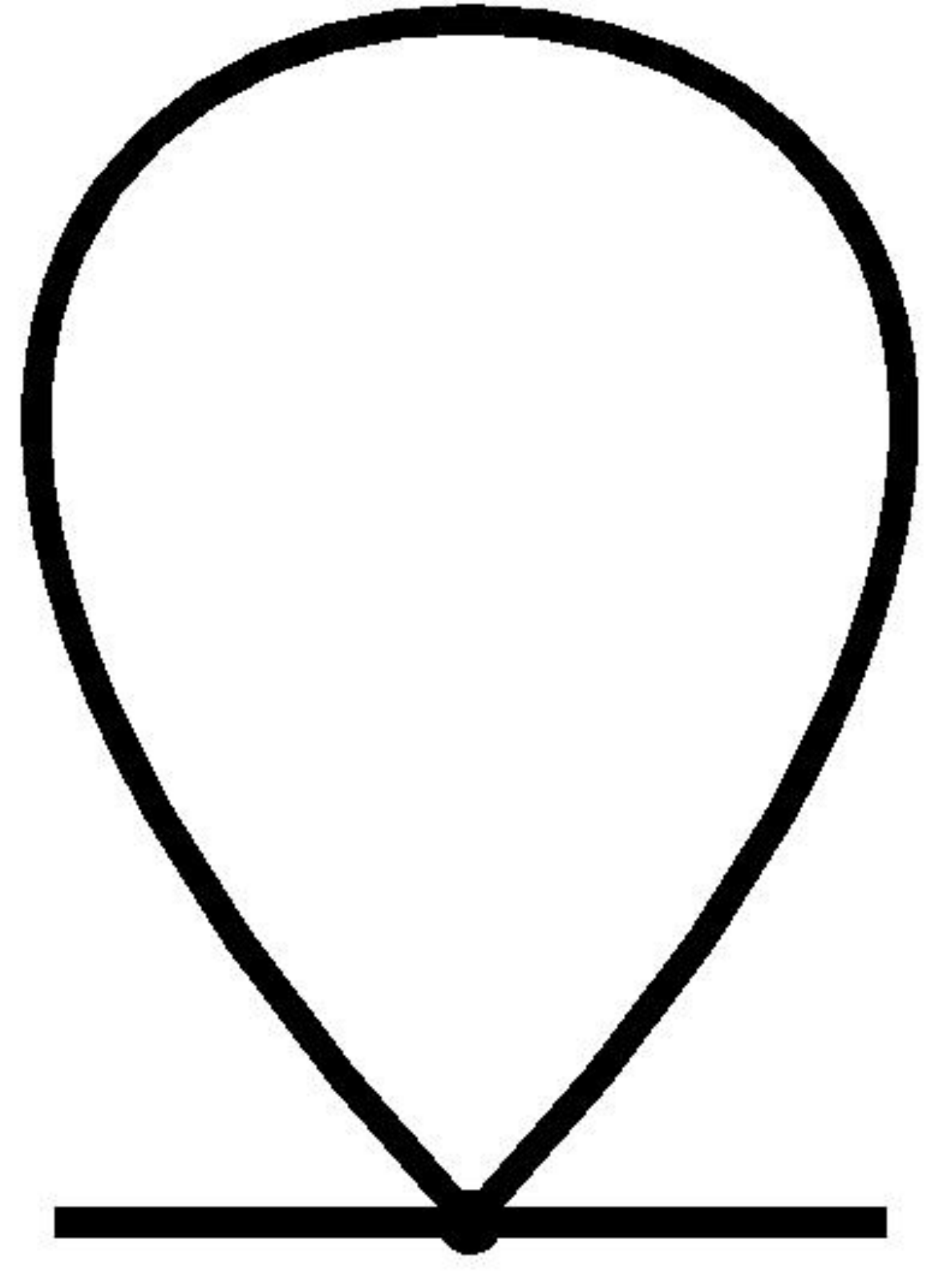}
   \label{fig:subfig1}
 }
 \subfigure[]{
   \includegraphics[height=1.8cm] {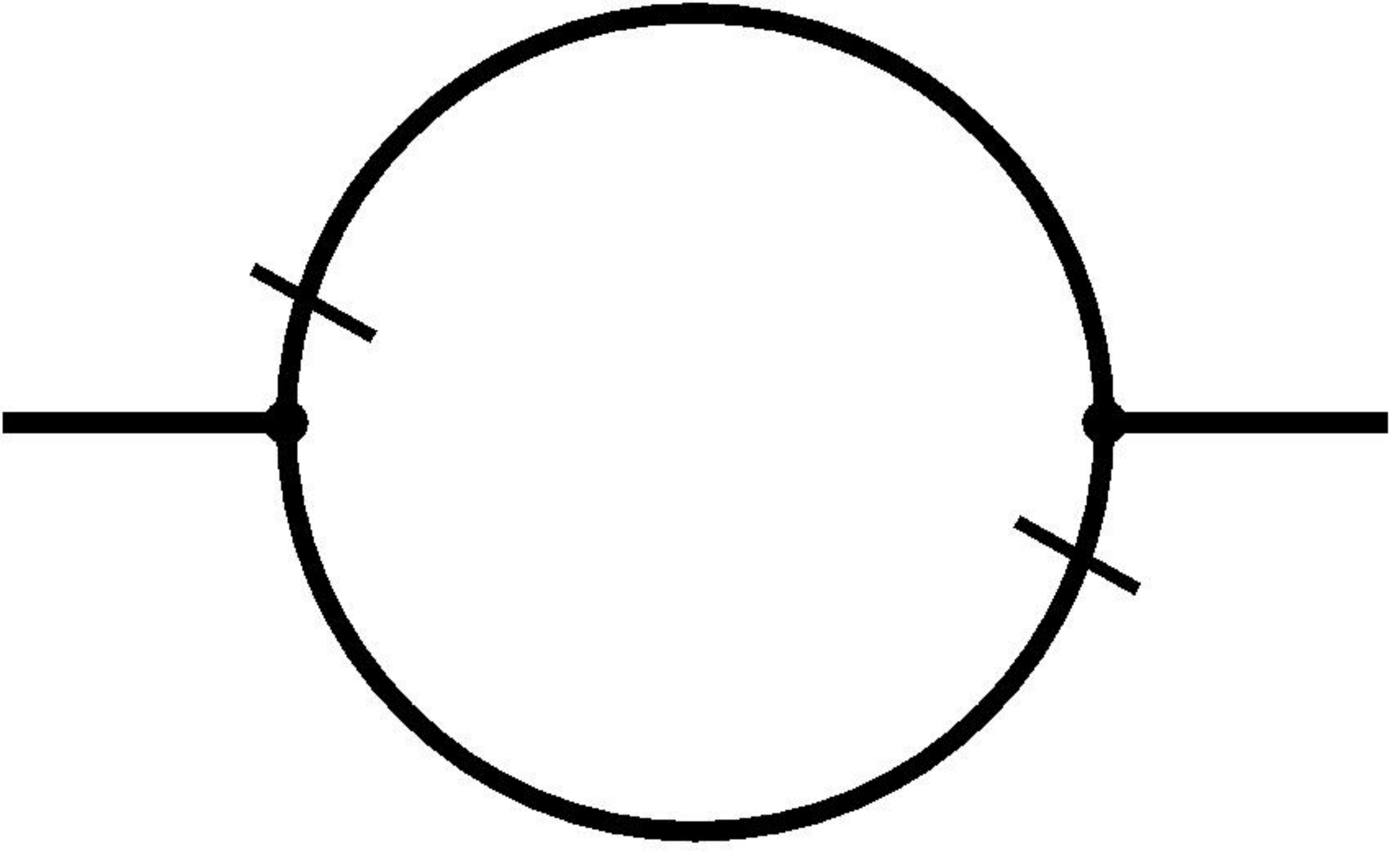}
   \label{fig:subfig2}
 }
  \subfigure[]{
   \includegraphics[height=1.8cm] {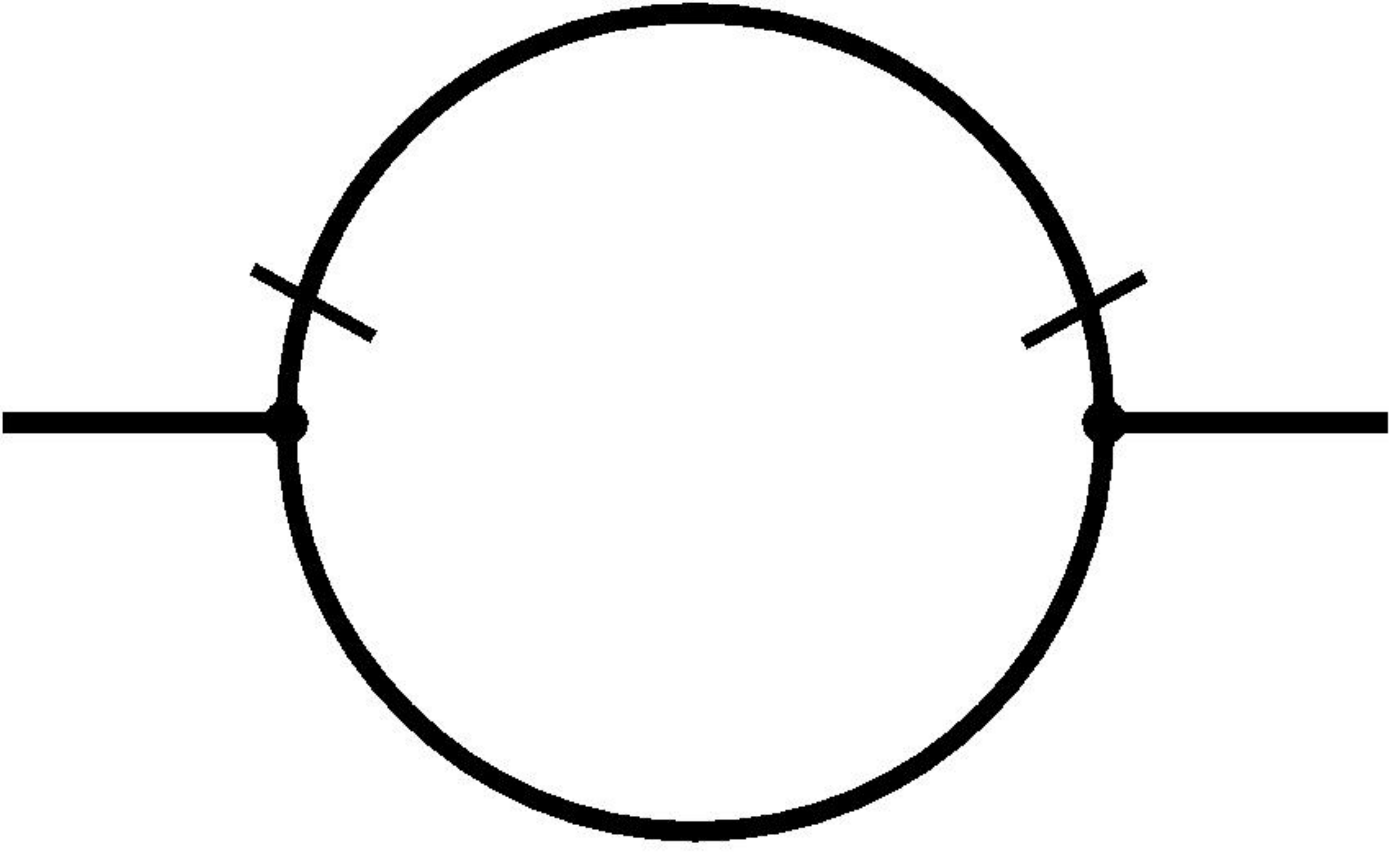}
   \label{fig:subfig3}
 }
 \subfigure[]{
   \includegraphics[height=1.8cm] {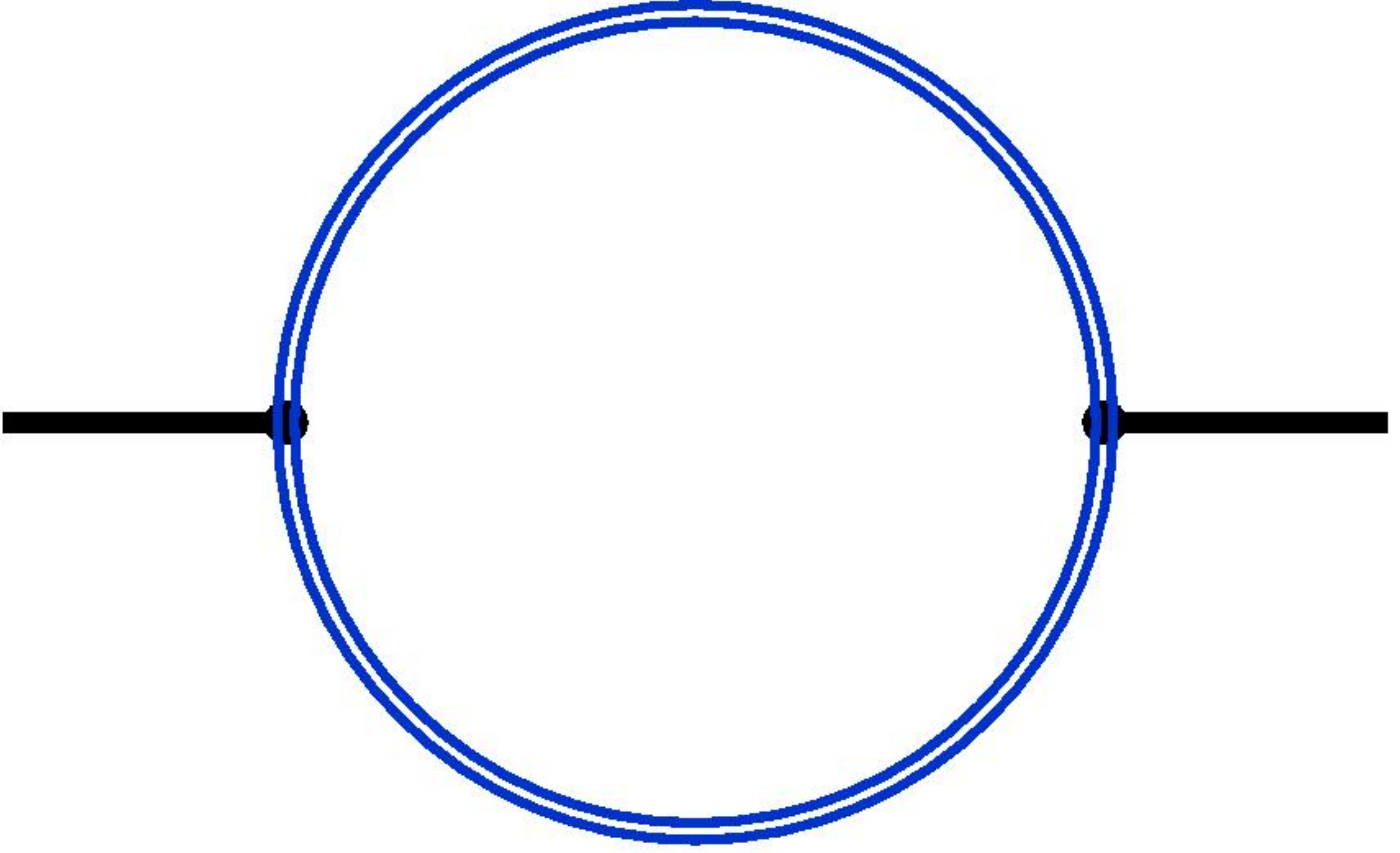}
   \label{fig:subfig4}
 }
\caption{\label{graphs}\small One loop contributions to the beta-function.
The dashed lines denote derivatives acting on propagators.}
\end{center}
\end{figure}

We first focus on the bosonic contribution. We are interested in those terms that produce a log-divergence in the effective action. Since the unique dimension-two operator  that can be constructed from the background currents is $K\wedge *K=K_\mu K^\mu $, we can set the background gauge field to zero (it appears in the effective action only through the field strength in operators of dimension four and higher). The possible diagrams that contribute to renormalization of $K^2$ are shown in fig.~\ref{fig:subfig1}-\ref{fig:subfig3}. The first diagram is the same as in the model without the WZ term, and its log-divergence can be taken from  \cite{Zarembo:2010sg}: 
\begin{equation}\label{I1a}
 I _{1a}=\frac{\ln\Lambda }{4\pi }\int_{}^{}\mathop{\mathrm{tr}}\nolimits_B
 \mathop{\mathrm{ad}}K\wedge *\mathop{\mathrm{ad}}K,
\end{equation}
where $\mathop{\mathrm{tr}}\nolimits_B$ denotes the trace over bosonic generators of $\mathfrak{g}$. The contribution of the diagrams \ref{fig:subfig2} and \ref{fig:subfig3} is proportional to
$$
 I_{1b,c} \sim \chi ^2\int_{}^{}\frac{d^2p}{\left(2\pi \right)^2}\,\,\frac{\varepsilon ^{\mu \lambda }\varepsilon ^{\nu \rho }p_\lambda p_\rho }{p^4}=
 -\chi ^2\eta ^{\mu \nu }\,\frac{\ln\Lambda }{4\pi }\,,
$$
where $\mu $, $\nu $ are 2d Lorentz indices that are contracted with $K_\mu K_\nu $. The full bosonic contribution is
\begin{equation}\label{effactionbos}
 I_{\rm bos}=\left(1-\chi ^2\right)\frac{\ln\Lambda }{4\pi }\int_{}^{}\mathop{\mathrm{tr}}\nolimits_B
 \mathop{\mathrm{ad}}K\wedge *\mathop{\mathrm{ad}}K.
\end{equation}

There is actually a simper way to arrive at the same result. Once the cross term is absorbed in the re-definition of the covariant derivative, as in (\ref{quadbosonic}), the effective action will depend on the field strength of the new gauge field $A+\chi *K$, but only through operators of dimension four or higher. The covariant derivative can thus be dropped altogether, and only the diagram \ref{fig:subfig1} will contribute to the beta-function, but now with the coefficient $(1-\chi ^2)$.

In order to evaluate the fermionic contribution, we need to fix the kappa-symmetry. Since the kappa-symmetry transformations act as linear shifts at the quadratic level, the gauge degrees of freedom simply drop from the action, but, as a result, the Dirac operator has a lot of zero modes which have to be excluded when computing the determinant of the Dirac operator.

From equation \eqref{kappa_condition} we can easily see that one can take:
\bee
\label{Kappaproj}
C^+\propto \gamma v &\text{and}& C^-\propto \bar\gamma v
\eee
where $v$ is some two dimensional vector and we have defined:
\bee\label{gammagammabar}
\gamma=\frac{1+\kappa\sigma_3+\chi \sigma_1}{2}	\nn\\
\bar \gamma=\frac{1-\kappa\sigma_3-\chi \sigma_1}{2}
\eee
for which, given the condition \eqref{kappathroughchi},  the following relations hold:
\be
\gamma^2=\gamma ,
\qquad \bar\gamma^2=\bar{\gamma },\qquad \gamma\bar\gamma=\bar\gamma\gamma=0.
\ee

Using these notations, we can rewrite the action \eqref{quadferm} as:
\bee
\label{quadferm+-}
 S_F^{(2)}
&=&\frac{1}{2}\int_{}^{}\mathop{\mathrm{Str}}
 X
 \left(
 -D_+\bar\gamma\mathop{\mathrm{ad}}K_-+ D_-\gamma\mathop{\mathrm{ad}}K_+
 \right. \nn\\&&\left.
 -\mathop{\mathrm{ad}}K_+ \sigma_1
\bar\gamma\mathop{\mathrm{ad}}K_-+ \mathop{\mathrm{ad}}K_-\sigma_1\gamma \mathop{\mathrm{ad}}K_+
 \right) X.
\eee
where now we have switched to the component notation. So, for example\footnote{We  use the $(-,+)$ signature for the worldsheet metric, which we also take to be flat Minkowski metric (the beta-function obviously does not depend on the 2d curvature).}, $D_+=D_0+D_1$, $D_-=D_0-D_1$, and similarly for $K_\pm$.

It is convenient to write the second order fermionic action in the following equivalent way:
\be
 S_F^{(2)}=\frac{1}{2}\int_{}^{}\mathop{\mathrm{Str}}
 X_I
 \left(
 -D_+\bar\gamma+ D_-\gamma-\mathop{\mathrm{ad}}K_+\sigma_1
\bar\gamma+ \mathop{\mathrm{ad}}K_-\sigma_1\gamma \right)\sigma_1\Gamma
 X_J
\ee
where
\be
\Gamma=\sigma_1(\bar\gamma\mathop{\mathrm{ad}}K_-+\gamma\mathop{\mathrm{ad}}K_+).
\ee
We thus easily see that the $X_I$'s  in the kernel of $\Gamma$ do not contribute to the action, and thus can be gauged away by imposing
\be
X_I=(\Gamma\,X)_I,
\ee
which is a gauge-fixing condition compatible with the kappa-symmetry projectors \eqref{Kappaproj}.

Finally the fermion contribution to the effective action is
\begin{equation}
 I _{\rm fer}=-\frac{1}{2}\,\mathop{\mathrm{tr}}\nolimits'\ln \left(\slashed D+M\right),
\end{equation}
where the Dirac operators is defined as
\be
\slashed D+M=-D_+\sigma_1\bar\gamma\sigma_1+ D_-\sigma_1\gamma\sigma_1-\mathop{\mathrm{ad}}K_+
\bar\gamma\sigma_1+ \mathop{\mathrm{ad}}K_-\gamma\sigma_1,
\ee
and $\mathop{\mathrm{tr}}'$ denotes the trace over the complement of $\Gamma $ in the fermionic part of the superalgebra $\mathfrak{g}$.

The log-divergence comes from the diagram in fig.~\ref{fig:subfig4}. Its contribution is given by
\begin{equation}
 I_{1d}= -\frac{1}{2}\int_{}^{}\frac{d^2p}{\left(2\pi \right)^2}\,\,
 \frac{p_+p_-}{p^4}\,\mathop{\mathrm{tr}}M\sigma _1\bar{\gamma }\sigma _1
 M\sigma _1\gamma \sigma _1
\end{equation}
 Taking into account that $\mathop{\mathrm{tr}}\bar{\gamma }\sigma _1\gamma \sigma _1=\kappa ^2$, we find:
\begin{equation}
 I_{\rm fer}=-\kappa ^2\frac{\ln\Lambda }{4\pi }\int_{}^{}\mathop{\mathrm{tr}}\nolimits_F
 \mathop{\mathrm{ad}}K\wedge *\mathop{\mathrm{ad}}K,
\end{equation}
where the trace is now taken over the fermion generators. Combining this with the bosonic contribution, we find the complete one-loop beta-function:
\be
I=\frac {\kappa^2}{4\pi}\,\ln \Lambda\int_{}^{}  \mathop{\mathrm{Str}}\nolimits_{\rm adj}K\wedge *K,
\ee
where $\mathop{\mathrm{Str}}=\mathop{\mathrm{tr}}_B-\mathop{\mathrm{tr}}_F$.  This $\beta$-function coincides with the one computed in \cite{Zarembo:2010sg} for the coset without the WZ term, up to an additional factor of $\kappa^2$.

The pure NSNS coset with $\kappa =0$, $\chi =1$ is a fixed point of the renormalization group, as expected.  For a generic coset with $\kappa \neq 0$, the beta-function is  proportional to the Killing form of the superalgebra. If the Killing form vanishes identically, so will the beta-function. This in particular happens for the superalgebras in the $\mathfrak{psu}(n|n)$ and $\mathfrak{osp}(2n+2|2n)$ series (see \cite{Zarembo:2010sg} for a systematic classification of possible conformal models). This means that the vanishing of the beta-function occurs for Super($AdS_3\times S^3$)  and Super($AdS_3\times S^3\times S^3$) cosets, keeping in mind that $D(2,1;\alpha)$ is a generalization of $OSp(4|2)$. We thus can see that the conditions that ensure conformal invariance are not modified by the introduction of the WZ term.
It should be also possible to reverse the logic of this section, not imposing \eqref{kappathroughchi} from the very beginning. The condition for conformal invariance then again leads to the same relation between the couplings.

\section{Mass spectrum in the light-cone gauge}

The background-field expansion can be also used to quantize the string in the light-cone gauge associated with a BMN geodesic. The background field $\bar{g}_{L,R}$ then corresponds to a point-like string moving along a light-like geodesic in the background geometry, say orbiting the sphere in the $AdS_3\times S^3$ geometry at the speed of light. As a specific example we will consider the $D(2,1;\alpha )^2/SU(1,1)\times SU(2)^2$ coset. The BMN geodesic then corresponds to taking the background field in the form
\begin{equation}
 \bar{g}_{L,R}=\,{\rm e}\,^{i(D+J)\tau},
\end{equation}
where $D$ is the dilatation generator in $\mathfrak{d}(2,1;\alpha )$, $J$ is a rotation generator, and $\tau $ is the worldsheet time. $D$ generates global time translations in $AdS_3$ and $J$ generates rotations of the three-sphere along some fixed axis. For the background currents we then get:
\begin{equation}
 A=0,\qquad K=i(D+J)d\tau .
\end{equation}
In particular, $K_\pm=i(D+J)$, will be denoted simply by $K$.

In the bosonic action (\ref{quadbosonic}), the covariant derivative contains a gauge connection $\chi K$, but since $K$ is a constant element of $\mathfrak{g}$, the gauge field can be absorbed by a field redefinition: $X_2\rightarrow \,{\rm e}\,^{-\chi *K}X_2\,{\rm e}\,^{\chi *K}$, after which (\ref{quadbosonic}) becomes canonically normalized action for a collection of free bosons with the mass matrix
\begin{equation}\label{MB}
 \mathcal{M}_B^2=-\kappa ^2\left(\mathop{\mathrm{ad}}K\right)^2.
\end{equation}

Let us move to consider the fermionic part. Starting from \eqref{quadferm+-} we can perform a rotation with a matrix $e^{-\frac i 2 s \sigma_2}$ with
$\cos s=\kappa$, $\sin s=\chi$,
that acts on the Pauli matrices as follows:
\bee
\sigma_1\rightarrow \kappa\sigma_1+\chi\sigma_3\nn\\
\sigma_3\rightarrow \kappa\sigma_3-\chi\sigma_1,
\eee
so the matrices defined in (\ref{gammagammabar}) become $\gamma \rightarrow (1+\sigma _3)/2$, $\bar{\gamma }\rightarrow (1-\sigma _3)/2$.

After the rotation we get:
\bee
S_F^{(2)}&=&\frac{1}{4}\int_{}^{}\mathop{\mathrm{Str}}
 X
 \left[
 -D_+(1-\sigma_3)+ D_-(1+\sigma_3)\right.\nn\\&&\left.-\mathop{\mathrm{ad}}K_+(\kappa\sigma_1+\chi\sigma_3)(1-\sigma_3)
+ \mathop{\mathrm{ad}}K_-(\kappa\sigma_1+\chi\sigma_3)(1+\sigma_3)\right]\Pi 
 X,\nn\\
\eee
where:
\be
\Pi =\frac{1-\sigma_3}{2}\mathop{\mathrm{ad}}K_-+\frac{1+\sigma_3}{2}\mathop{\mathrm{ad}}K_+=\mathop{\mathrm{ad}}K.
\ee
Using $K_+=K_-=K$, $D_\pm=\partial _\pm$ we find:
\be
S_F^{(2)}=\frac{1}{4}\int_{}^{}\mathop{\mathrm{Str}}
 X
\left\{
 -\partial _+(1-\sigma_3)+ \partial _-(1+\sigma_3)
 +2(\chi-i\kappa \sigma _2)\mathop{\mathrm{ad}}K\right\}
[K, X].
\ee
The $\chi $-dependence can be absorbed by a gauge transformation:
\begin{equation}\label{gauge-sigma}
 X\rightarrow \,{\rm e}\,^{i\chi \left(D+J\right)\sigma }X
 \,{\rm e}\,^{-i\chi \left(D+J\right)\sigma },
\end{equation}
under which $\partial _\pm\rightarrow \partial _\pm\pm\chi \mathop{\mathrm{ad}}K$. Since $*K=-i(D+J)d\sigma $, this is the same gauge transformation that eliminates the cross term in the bosonic action (\ref{quadbosonic}). It should be noted that these transformations potentially change boundary conditions, because the parameter of the transformation non-trivially depends on $\sigma $.

After this transformation the Lagrangian can be written in the standard 2d Dirac form:
\begin{equation}
 \mathcal{L}^{(2)}_F=\frac {1}{2}\mathop{\mathrm{Str}}\bar{\psi }\left(\rho ^\mu \partial _\mu -\mathcal{M}_F\right)\psi ,
\end{equation}
where
\begin{equation}
 \psi =\begin{pmatrix}
  X_1  \\
   X_3 \\
 \end{pmatrix},
\end{equation}
$\rho ^\mu =(i\sigma _1,\sigma _2)$ and $\bar{\psi }=C\psi ^t$. The charge conjugation is defined by the action of the following operator:
\begin{equation}
 C=-\sigma _2\mathop{\mathrm{ad}}K.
\end{equation}
 In the chosen basis of Dirac matrices, $\rho ^\mu {}^tC=-C\rho ^\mu $ indeed holds. It is understood that those components of $X_1,X_3$ that commute with $K$ (the zero eigenvectors of $C$) are eliminated by the kappa-symmetry gauge choice.

The mass matrix of the fermions is of the form:
\begin{equation}\label{MF}
 \mathcal{M}_F=i\kappa \mathop{\mathrm{ad}}K.
\end{equation}
From this equation, as well as from analogous equation for bosons $(\ref{MB})$, we see that  the mass spectrum is determined by the eigenvalues of $i\mathop{\mathrm{ad}}K$ multiplied by a factor of $\kappa $. In particular, at $\kappa =0$ all the fluctuation modes become massless.

As an example, we compute the BMN spectrum of the $AdS_3\times S^3\times S^3$ sigma model with mixed RR/NSNS fluxes \cite{Berenstein:2002jq}. The underlying symmetry algebra actually does not depend on the fluxes, and the effect of the NSNS flux is just an overall multiplication of the mass eigenvalues by\footnote{Here we are only talking about the spectrum. It is unlikely that the effect of $\kappa \neq 1$ on the interactions between the BMN modes can be reduced to simple rescalings.} $\kappa $ and modification of the boundary conditions due the gauge transformation (\ref{gauge-sigma}) which was necessary to bring the kinetic terms in the action to the canonical form. To make the discussion self-contained we re-derive the spectrum using the commutation relations of the $\mathfrak{d}(2,1;\alpha )$ superalgebra, as in \cite{Babichenko:2009dk}.

 We denote the bosonic generators of the $\mathfrak{d}(2,1;\alpha )$ by $S_\mu $, $L_n$ and $R_{\dot{n}}$. They form three copies of  $\mathfrak{sl}(2)$  and are normalized as
 \begin{equation}\label{norm}
  \mathop{\mathrm{Str}}S_{\mu}S_\nu =\frac{1}{4}\,\eta_{\mu\nu},
  \qquad 
  \mathop{\mathrm{Str}}L_nL_m=\frac{1}{4\cos^2\phi}\,\delta_{nm},
  \qquad 
  \mathop{\mathrm{Str}}R_{\dot{n}}R_{\dot{m}}=
  \frac{1}{4\sin^2\phi}\,\delta_{\dot{n}\dot{m}},
 \end{equation} 
where $\eta_{\mu\nu}$ is the metric of the $(++-)$ signature, and the angle $\phi$ is related to the $\alpha$ of the $\mathfrak{d}(2,1;\alpha )$ by $\alpha=\cos^2\phi$. The supercharges of $\mathfrak{d}(2,1;\alpha)$ are in the tri-spinor representation of $\mathfrak{sl}(2)^3$: $Q_{ a \alpha\dot{\alpha}}$.
 The continuous parameter $\cos^2\phi$ appears only in the anticommutator of the supercharges and in the norm (\ref{norm}). The rest of the commutation relations are fixed by the $\mathfrak{sl}(2)^3$ symmetry.
 
While the dilatation generator is a non-compact element of the first $\mathfrak{sl}(2)$, defined more or less unambiguously: $D=S_3$, there is  a certain freedom in defining the rotation generator. {\it A priori} $J$ can be an arbitrary linear combination of the Cartan generators of the compact $\mathfrak{sl}(2)$'s: $J=C_1L_3+C_2R_{\dot{3}}$. However, the light-cone condition requires that $D+J$ is a null element of the superalgebra, which imposes the constraint $C_1^2/\cos^2\phi+C_2^2/\sin^2\phi=1$. We also want the ground state of the string to be a BPS state, or, in other words, the classical solution should preserve maximally possible amount of supersymmetry. The unbroken supersymmetries correspond to the supercharges that commute with $D+J$. Since $i[D+J,Q_{a\alpha\dot{\alpha}}]=(\pm 1\pm C_1\pm C_2)Q_{a\alpha\dot{\alpha}} $ with eight possible combinations of signs, the supersymmetry is preserved when $C_1\pm C_2=\pm 1$. This, together with the zero norm condition fixes, up to a sign, $C_1=\cos^2\phi$, $C_2=\sin^2\phi$ and thus
\begin{equation}\label{D+J}
  D+J=S_3+\cos^2\phi\,L_3+\sin^2\phi\,R_{\dot{3}}\,.
\end{equation}
The eigenvalues of the adjoint action of this operator are $\pm 1$, $\pm\cos^2\phi$, $\pm\sin^2\phi$ on the bosonic subalgebra and $s_1+s_2\cos^2\phi+s_3\sin^2\phi$, with $s_i=\pm 1/2$, on the fermionic generators. There is an additional zero eigenvalue in the bosonic sector, which originates from a linear combination of $L_3$ and $R_{\dot{3}}$ orthogonal to (\ref{D+J}). From (\ref{MB}), (\ref{MF}) we then get the mass spectrum of the string modes:
\begin{eqnarray}
 M^2_B&=&\left\{
 \kappa^2 ,\kappa^2 ,\kappa^2\cos^4\phi ,\kappa ^2\cos^4\phi ,
 \kappa ^2\sin^4\phi ,\kappa ^2\sin^4\phi ,0 
 \right\}
\nonumber \\
M_F&=&\left\{
 \kappa ,-\kappa ,\kappa \cos^2\phi ,-\kappa \cos^2\phi ,
 \kappa \sin^2\phi ,-\kappa \sin^2\phi 
\right\}.
\end{eqnarray}
This cannot be a complete spectrum of a consistent string theory in ten dimensions. Additional massless modes (one massless boson and two massless fermions) come from the orthogonal $S^1$ direction of the $AdS_3\times S^3\times S^3\times S^1$ background, and is not described by the coset sigma-model. Interactions between coset and non-coset modes (at $\kappa =1$) are discussed in detail in \cite{Rughoonauth:2012qd,Sundin:2012gc}. 

\section{Discussion}

Let us first make few remarks of technical nature. The mildly non-local nature of the WZ term requires that the coefficient in front is quantized, in order to make the path integral (\ref{path_int}) independent of the three-dimensional continuation of the fields in the sigma-model to the interior of the three-dimensional domain $\mathcal{B}$ used to define the WZ action \cite{Witten:1983tw,Witten:1983ar}. With the normalization as in (\ref{mainaction}), (\ref{path_int}), and assuming that the supertrace reduces to the usual trace for the bosonic generators, the quantization condition becomes
\begin{equation}
 4\chi \sqrt{\lambda }=k
\end{equation}
with $k$ integer.

The action (\ref{mainaction}) with $G=D(2,1;\alpha )$, and supplemented with one free boson, should describe the GS superstring on $AdS_3\times S^3\times S^3\times S^1$ supported by a combination of the NSNS and RR three-fluxes. When $\alpha =0$, or if we just start with $G=PSU(1,1|2)$ and add four free bosons, the background degenerates to $AdS_3\times S^3\times T^4$. Although we have not checked that the GS action reduces to the coset action upon fixing the kappa-symmetry gauge, this more or less follows from symmetries and the results of \cite{Babichenko:2009dk}, where the equivalence between the coset action and the standard  GS action in curved space \cite{Cvetic:1999zs} was demonstrated in the absence of the WZ term. The WZ term in the string action and the B-field in the supergravity have the same symmetries. The condition for vanishing of the beta-function should be equivalent to the supergravity equations of motion. One could in principle directly compare  the coset action to the GS action from \cite{Cvetic:1999zs} by using the background field expansion from sec.~\ref{sec:background}.

We have shown that the semi-symmetric permutation cosets with the WZ term remain integrable. This generalizes classical integrability of the bosonic principal chiral field with the WZ term  \cite{TakhtajanVeselov84}. For the supersymmetric cosets, the conditions for integrability, kappa-symmetry and conformal invariance (zero beta-function) turn out to be equivalent and require a specific relationship between the couplings of the GS and WZ terms.

The pure RR $AdS_3$ backgrounds thus admit a number of integrable deformations which in addition to switching the B-field include squashing of the three-sphere \cite{Orlando:2012hu}. It would be interesting to work out the consequences of integrability for all these backgrounds by generalizing known results for $AdS_3\times S^3\times T^4$ and $AdS_3\times S^3\times S^3\times S^1$ supported by pure RR flux. This includes the algebraic curve construction for quasi-periodic classical solutions, the Bethe ansatz equations for the quantum spectrum, and the Y-system that takes into account finite-size effects. Of particular interest is the limit $\chi \rightarrow 1$, which may provide a link between AdS/CFT integrability and representation-theory based methods of the worldsheet CFT \cite{Maldacena:2000hw,Maldacena:2000kv,Maldacena:2001km}.
Generalizing classification of integrable boundary conditions for the $\mathbbm{Z}_4$ cosets \cite{Dekel:2011ja} to include the WZ coupling should be really helpful in this respect, as it may provide a direct link to the boundary conformal perturbation theory of \cite{Quella:2007sg}.

\subsection*{Acknowledgments}
We are grateful to A.~Babichenko for drawing our attention to this problem and for many useful discussions.
We would like to thank G.~Arutyunov, O.~Ohlsson~Sax, A.~Sfondrini, D.~Skinner, D.~Sorokin, A.~Tseytlin and P.~Vieira for discussions and suggestions, and to A.~Babichenko and A.~Torrielli for comments on the draft.
K.Z. would like to thank the Perimeter Institute for kind hospitality during the course of this work.  A.C. would like to thank the "Fondazione Angelo Della Riccia" for the financial support during her staying at Nordita Institute.
The work of K.Z. was supported in part by  the RFFI grant 10-02-01315, and in part
by the Ministry of Education and Science of the Russian Federation
under contract 14.740.11.0347.
This work was partially done under the Padova University Research Grant 
CPDA119349

\appendix

\bibliographystyle{nb}

\end{document}